\documentclass{JHEP3}

\usepackage{amsmath,amsfonts,amssymb,amsthm,amstext,eucal}
\usepackage{graphicx,lscape}
\usepackage{epsfig}

\newcommand{\bea}{\begin{eqnarray}} 
\newcommand{\eea}{\end{eqnarray}}
\def\beann{\begin{eqnarray*}} 
\def\eeann{\end{eqnarray*}}

\newcommand{\beq}{\begin{equation}}
\newcommand{\eeq}{\end{equation}}

\newcommand{\ba}{\begin{array}} 
\newcommand{\ea}{\end{array}}

\def\ben{\begin{enumerate}} 
\def\een{\end{enumerate}}

\def\4{\tilde }
\def\5{\bar }  
\def\6{\partial } 
\def\7{\hat }

\font\mybb=msbm10 at 10pt
\def\bb#1{\hbox{\mybb#1}}
\def\bR {\bb{R}}
\def\bZ {\bb{Z}}

\def\bC {\bb{C}}

\newcommand{\SO}{\mathrm{SO}}
\newcommand{\SL}{\mathrm{SL}}

\preprint{hep-th/0309199 \\
WIS/23/03-SEPT-DPP\\
UPR-1052-T}
\title{G\"odel-type Universes and the Landau problem}
\author{Nadav Drukker\footnote{Work done mostly while still at the 
Weizmann Institute}\\
The Niels Bohr Institute, Copenhagen University\\
Blegdamsvej 17, DK-2100 Copenhagen, Denmark.\\
\email{drukker@nbi.dk}}
\author{Bartomeu Fiol\\
The Weizmann Institute of Science, Department of Particle Physics\\
Herzl Street 2, 76100 Rehovot, Israel\\
\email{fiol@weizmann.ac.il}}
\author{Joan Sim\'{o}n$^*$\\
Department of Physics and Astronomy, David Rittenhouse Laboratory, University 
of Pennsylvania,\\
209 South 33rd Street, Philadelphia PA 19104-6396, USA.\\
\email{jsimon@bokchoy.hep.upenn.edu}
}

\abstract{We point out a close relation between a family of G\"odel-type 
solutions of 3+1 General Relativity and the Landau problem in $S^2,\bR^2$ and
$\bb{H} _2$; in particular, the classical geodesics correspond to Larmor 
orbits in the Landau problem. We discuss the extent of this relation, by
analyzing the solutions of the Klein-Gordon equation in these 
backgrounds. For the $\bR^2$ case, this relation was independently 
noticed in hep-th/0306148. Guided by the analogy with the Landau problem, we 
speculate on the possible holographic description of a single chronologically 
safe region.}


\begin{document}

\section{Introduction.}

Renewed interest in spacetimes with closed timelike curves has been sparked
in part by the work of Gauntlett et al. \cite {gauntlett}, who describe 
all SUSY solutions of ${\cal N}=1$ 5d supergravity with a 
timelike Killing vector, and note that the generic solution has closed timelike
curves (CTCs). 
One of the surprises of this analysis was the discovery of a maximally 
supersymmetric 5d relative of the original 4d G\"odel solution \cite{godel}. 
Similar techniques have been applied to the classification of 
SUSY solutions of 11d supergravity \cite{pakis} with a timelike Killing 
vector, and again, solutions with closed timelike curves are commonplace. 
In particular, SUSY G\"odel type metrics were also found.

The existence of G\"odel type solutions of 10d and 11d supergravity 
raises the issue of whether these are valid backgrounds in string/M 
theory\footnote{To the best of our knowledge, the first appearance 
of a 
G\"odel type solution in the string theory literature was in
\cite{horowitz}, as the KK reduction of a 5d chiral model, although the 
interpretation as a G\"odel-type metric was not explicitly mentioned there.}, a
question addressed in a number of recent works \cite{herdeiro, horava, 
harmark, we, rey, moshe, dan, brace, taka}. The validity of other supergravity 
solutions with closed timelike curves has been discussed in \cite{gibbons, 
specprop, sabra, emparan, dyson}. 

In particular, in our previous work \cite{we}, we argued that a class of 
G\"odel type solutions of type II supergravity are not, despite $g_s=0$ 
appearances, valid backgrounds in string theory. We argued for this by 
showing that a specific D-brane probe, a supertube \cite{mateos}, develops 
negative mass modes in this background,
following a similar computation performed in \cite{emparan}. Since the 
worldvolume dynamics of D-branes captures the spacetime where they live, we 
interpreted this sickness as a problem of the spacetime itself, not just of the
probe.

Nevertheless, there are many questions unanswered concerning G\"odel-type 
universes. In this respect, the authors of \cite {horava} made an intriguing 
observation for these spacetimes: If one applies Bousso's prescription 
\cite{bousso} for constructing holographic screens (see \cite{holorev} for a 
review) to these solutions, one finds that starting at any point in spacetime, 
the screen is inside the corresponding chronologically safe region. The 
precise meaning of these holographic screens when they are not at asymptotia
has yet to be fully clarified, but the take of \cite{horava} seems to be that
in a fully quantum gravity theory on these backgrounds, any individual observer
has a unitary description of the physics, encoded in the holographic 
screen.

For the reasons mentioned above, the G\"odel type metric is not a valid 
solution of string theory; however, the question of holography for this
background can be rephrased as follows: in \cite{we}, we presented a new 
solution that locally is a G\"odel type space, but at the radius where closed 
timelike curves would start 
to appear, we placed a domain wall made out of supertubes, such that the 
overall solution is free of CTCs. This domain wall clearly breaks the 
translation invariance of the original solution. Bousso's prescription 
still yields the same holographic screen as in \cite{horava} for an observer
at the origin of this new solution, and the question of which are the degrees
of freedom at the holographic screen, still remains. 

The possibility of having holographic screens that are not at asymptotia
raises many questions that were not present in the more familiar case of 
$AdS$/CFT holography. A very 
first question is what are the bulk (string/gravity) degrees of freedom that
one should associate to a particular screen. It is worth noting that even
when the metric can be embedded in a string theory solution, and the string 
spectrum computed at $g_s=0$ \cite{russo, russohet, harmark}, the free string 
spectrum 
shows no immediate hint of a holographic description, or the possibility 
of a compact holographic screen. For relevant spacetimes as de Sitter and 
FRW cosmologies, Bousso's prescription yields observer dependent 
holographic screens (although in the case of de Sitter 
it is also possible to place the screen at asymptotia \cite{bousso}), 
so it would be clearly desirable to obtain concrete 
examples of observer dependent holography.

In the present work we will be interested in a family of solutions of 3+1 
General Relativity, that extend G\"odel's original solution 
\cite{godel}. 
That solution corresponds to a homogeneous spacetime, with rigid rotation, 
negative cosmological constant and a perfect fluid as source. G\"odel's 
original solution was generalized, among others, by Rebou\c{c}as and 
Tiomno \cite{tiomno}, who found all the spacetime homogeneous solutions of 
General Relativity with 
rigid rotation. Their solutions typically have non-zero cosmological constant
and non-vanishing stress energy tensor, and the metric in polar coordinates 
can be written as
\bea
ds^2=-\left(dt+\frac{\Omega}{l^2} \sinh^2 lr d\phi\right)^2+ 
\frac{\sinh^2\; 2lr}{4l^2}d\phi^2 +dr^2+dx_3^2\,.
\label{eq:rebou}
\eea

The main observation of the present paper is that the physics in these 
metrics have a close relationship with the problem of a charged particle in 
$\bb{H}_2$, $\bR^2$ or $S^2$---depending on the 
sign of $l^2$---coupled to a magnetic field of strength given by $\Omega$ and
an effective charge given by the energy (the conserved quantity associated to
the time isometry). For the $l^2\rightarrow 0$ case, this similarity was 
independently noticed and discussed in \cite{rey}. As we will see, the 
presence of closed timelike curves in these metrics translates into the 
absence of non-periodic orbits in the analogous
particle problem. G. Gibbons has suggested \cite{pgibbons} that this might be
a more general phenomenon.

At the classical level, the analogy between the G\"odel-type solutions and
the Landau problem has a first manifestation in the fact, shown in section
2, that all the geodesics project to circles, in the surface with constant $t$
and constant $x_3$. In particular, the projection of some spacelike 
geodesics are closed timelike curves. These projections of course are 
reminiscent of 
the Larmor orbits for an electron moving in a magnetic field. At the quantum 
level, we analyze in section 3 the solutions to the Klein-Gordon equation in 
these backgrounds (first studied in \cite{figue}), and note that the 
wavefunctions closely resemble those of the Quantum Hall Effect (QHE) on
$S^2,\bR^2$ and $\bb{H}_2$ respectively. The main difference is that there is
a rescaling in the wavefunctions, that causes a particular subset of them to
have most of their support within a region free of closed timelike curves. 
This leads to the suggestion, discussed in section 4, that when we 
restrict to a single chronologically safe region (e.g. by considering our 
solution \cite {we} with a domain wall) the relevant modes should be similar 
to the wavefunctions supported mostly inside it. We further argue that taking
into account gravity imposes a high energy cut-off in the bulk spectrum. We
then assume the existence of a holographic description of this single 
chronologically safe region, and deduce that the number of boundary degrees
of freedom scales linearly with the cut-off in the energy level. We conclude
presenting a simple model---based on considering a fuzzy version of the 
holographic screen---that reproduces this scaling.

\bigskip
\noindent
{\bf Note added:} As this manuscript was near completion, we learned from
G. Gibbons that some of the results presented in sections 2 and 3 were
independently obtained by himself and C. Herdeiro in unpublished work, as
mentioned in \cite{gibbons}.

\bigskip
\noindent
{\bf Note added:} After the first version of this work appeared, we learned 
from S.J. Rey that the analogy between G\"odel type metrics with $l^2=0$ with 
the Landau problem in the plane, was already noticed and discussed in a recent
paper by Y. Hikida and S.J. Rey \cite{rey}. Furthermore, the authors of 
\cite{rey} consider supergravity 
solutions with $l^2=0$ G\"odel type metrics and background field strengths 
turned on, which allows them to discuss also the geodesics and wavefunctions 
of particles charged under those fields. This introduces a qualitatively new
phenomenon, as compared to the discussion in the present paper, since some of
these charged geodesics can go beyond the radius of the chronologically safe
region. We briefly consider the possible implications of this observation in 
the last section.

\section{A family of G\"odel-type solutions} 

We will study a family of 3+1 metrics discussed by Rebou\c{c}as and Tiomno 
\cite{tiomno} that generalizes the original solution due to G\"odel. These 
are spacetime homogeneous metrics, with rigid rotation, and are characterized 
by
two parameters $(l,\Omega)$, both with the dimensions of inverse length. As in 
the original G\"odel solution, these spacetimes are of the form $\cal M_3 
\times \bR$, where ${\cal M}$ has signature $(2,1)$ 
and $\bR$ is a spatial direction that mostly plays no role in the discussion.
These solutions in general have non-zero cosmological constant and 
stress-energy tensor; if we use as sources a combination of perfect fluid, 
$U(1)$ gauge fields and/or a massless scalar, one can cover the range 
\cite{tiomno}
$-\infty < l^2 \leq \Omega^2$. The solutions with $l^2> \Omega ^2$ can be
realized \cite{olive}, if we allow for solutions with torsion; we will touch
upon these metrics only briefly.

It is useful to write the solution in different coordinate systems. The 
cylindrical symmetry of this family of solutions is manifest in polar 
coordinates
\bea
  ds^2=-\left(dt+\Omega \frac{\sinh^2 \;lr}{l^2} d\phi\right)^2+ 
  \frac{\sinh^2\;2lr}{4l^2}d\phi^2 +dr^2+dx_3^2\,,
 \label{eq:polar}
\eea
which will also be convenient when discussing the causal 
structure of these solutions, and the location of the holographic screens.
We can also rewrite the metric in Cartesian coordinates \cite{tiomno}
\bea
ds^2=-\left(dt+\frac{\Omega}{\sqrt{2}l}e^{2lx}dy\right)^2
+\frac{1}{2}e^{4lx}dy^2 +dx^2+dx_3^2\,.
\eea
Finally, for $l^2>0$, if we further make the change of coordinates
\bea
2lx=- \ln2lY\,,\hskip1cm y=\sqrt{2}X\,,
\eea
we obtain the metric in hyperbolic coordinates
\bea
ds^2=-\left(dt+\frac{\Omega}{2l^2}\frac {dX}{Y}\right)^2
+\frac{1}{(2lY)^2}\left(dX^2+dY^2\right)+
dx_3^2\,.
\eea
For some particular values of the parameters, this solution reduces to 
well-known ones. For $l^2=\Omega ^2/2$ we recover the G\"odel solution. 
For $l^2=\Omega ^2$ we get $AdS_3\times \bR$ \cite {rooman}. 

A relevant feature of these metrics is that they have closed timelike curves.
This is easiest to establish in polar coordinates, by considering curves of
constant $t,r,x_3$. The metric component $g_{\phi \phi}$ is
\bea
g_{\phi \phi}=\frac{\sinh^2 2lr}{4l^2}-\frac{\Omega ^2}{l^4}
\sinh^4 lr\,,
\eea
and changes sign at the radius
\bea
\tanh lr_c=\frac{l}{\Omega}\,.
\eea
Outside this radius $r_c$, there are closed timelike curves around the origin.
Therefore there are closed timelike curves for the range $l^2<\Omega ^2$.
We call the region of $r\leq r_c$ the chronologically safe cavity. Although 
for this discussion we fixed the origin, by the homogeneity of the metric
it is clear that exactly the same picture must hold at any point in space-time:
there is a chronologically safe cylinder of radius $r_c$ and closed timelike
curves outside it.

As we will see below, certain features of the solutions depend markedly on the
sign of $l^2$, so we briefly outline the different possibilities.

\noindent
\paragraph{Sphere $(l^2<0)$:} It is convenient to introduce the
new coordinates $R=i/2l$ and $\theta=r/R$. The metric \eqref{eq:rebou} becomes
\bea
ds^2=-\left(dt+4\Omega R^2 \sin^2\frac{\theta}{2} d\phi\right)^2+
R^2\left(d\theta^2+\sin^2 \theta d\phi^2\right)+dx_3^2\,.
\eea
Therefore, the 2+1 manifold ${\cal M}$ is a real line bundle over $S^2$.

\noindent
\paragraph{Flat plane $(l^2=0)$:} If we take the limit $l\to 0$ in the 
metric \eqref{eq:rebou}, we get a geometry originally obtained by Som and 
Raychaudhuri \cite{somray}, in which ${\cal M}$ is now a real bundle over 
$\bR^2$
\bea
ds^2=-\left(dt+\Omega r^2d\phi\right)^2+r^2d\phi^2+dr^2+dx_3^2\,.
\eea
This metric has appeared a number of times in solutions of string 
theory \cite{horowitz, russo, russohet} 
and more recently in \cite{horava, harmark}, where it was reinterpreted as a 
G\"odel-type solution in string theory.

\noindent
\paragraph{Hyperbolic plane $(l^2>0)$:} In this case, the manifold ${\cal M}$ 
is a real line bundle over $\bb{H}_2$. The range $0<l^2\leq \Omega ^2/2$ can 
be realized with the stress energy tensor of a perfect fluid and/or a $U(1)$ 
gauge field, whereas the range $\Omega ^2/2<l^2\leq \Omega^2$ can be realized 
by adding a scalar field to the sources \cite{tiomno}.

After these preliminary remarks, we come to an important observation: all 
those
metrics are of the generic form
\bea
ds^2=-(dt+A_i(x)dx^i)^2+h_{ij}(x)dx^idx^j\,,
\label{eq:sagnac}
\eea
where $x^i$ denote all the coordinates that are not $t$. For these $D$ 
dimensional metrics \eqref{eq:sagnac} the geodesic equation, in affine 
parametrization, is identical to the equations of motion of a charged 
particle moving in a $D-1$ surface with metric $h_{ij}$ coupled to
a magnetic field with gauge potential one-form given by $A(x) =A_i(x)\,dx^i$. 
Furthermore, the charge of the particle (or coupling) in this analogous 
system, is given by $\Pi_t$, the conjugate momentum to the timelike coordinate 
$t$. The metrics \eqref{eq:rebou} considered above correspond to the 
particular cases when, apart from the trivial extra direction $x_3$, the 
aforementioned surfaces have constant curvature ($S^2, \bR^2$ and $\bb{H}_2$) 
and the magnetic field is constant, i.e. proportional to the volume form. 
In light of this remark, we anticipate that there 
will be many similarities between physics in these spacetimes and the Landau 
problem in surfaces of constant curvature, which is reviewed briefly in the 
appendix.

\subsection{Isometries}

The spacetimes considered in this paper are homogeneous and have five 
independent Killing vectors \cite{tiomno}. In Cartesian coordinates, these 
are given by
\begin{equation}
  \begin{gathered}
    K_0 =\frac{\partial}{\partial t}\,,\qquad 
    K_1 =\frac{\partial}{\partial x_3}\,,\qquad 
    K_2 =\frac{\partial}{\partial y}\,, \\
    K_3  = -2ly\frac{\partial}{\partial y}+\frac{\partial}{\partial x}\,, \\
    K_4  = \frac{\Omega}{\sqrt{2}l^2}e^{-2lx}\frac{\partial}{\partial t}
   -y\frac{\partial}{\partial x}+\left(ly^2-\frac{1}{2l}e^{-4lx}\right)
   \frac{\partial}{\partial y}\,.
  \end{gathered}
\end{equation}
They satisfy the commutation relations
\bea
  & [K_0,K_r]=[K_1,K_r] = 0\,, \qquad r=0,\dots,4\,, & \\
  & [K_2,K_3]= -2lK_2\,, \qquad [K_2,K_4] = -K_3\,, 
    \qquad [K_3,K_4]=-2lK_4 \,.&
\eea
Notice that the subset $\{K_2,K_3,K_4\}$ form an $su(2),h_2,sl(2,R)$ 
subalgebra depending on the sign of $l^2$. It was checked in \cite{rooman} 
that when \eqref{eq:rebou} describes $AdS_3\times \bR$, that is when 
$l^2=\Omega^2$, the isometry group contains a second $\SL(2,\bR)$ factor, thus 
matching $\SO(2,2)$. 

\subsection{Geodesics}

The connection of the physics in these spaces with the Landau problem has a 
first manifestation when we consider the classical geodesics. As we will show, 
the projection of any geodesic on the $(r,\phi)$ plane is a circle, which is 
of course reminiscent of the Larmor orbits of an electron in a constant 
magnetic field. This fact was noted for the original G\"odel solution in 
\cite{pfarr}, and for the Som-Raychaudhuri solution in \cite{paiva}. The 
geodesics of higher dimensional versions of the $l^2=0$ case were discussed
recently in \cite{rey} where the analogy with Larmor orbits of the Landau 
problem is also pointed out. For the full family of solutions we are 
considering, the geodesics were qualitatively studied in \cite{calvao}, using 
polar coordinates. However, in that choice of coordinates, it is not 
manifest that the geodesics project to circles with arbitrary centers in the 
$(r,\phi)$ plane. We shall prove this statement in hyperbolic coordinates.

\noindent
\paragraph{Flat plane $(l^2=0)$:} The analysis of geodesics is easiest if 
we rewrite the metric in Cartesian coordinates
\bea
ds^2=-\left(dt+\Omega(ydx-xdy)\right)^2+dx^2+dy^2+dx_3^2 \,.
\eea
Let's denote by $\Pi_t$ and $\Pi_3$ the conserved quantities associated to
the $\partial_t$ and $\partial_3$ isometries. The geodesics satisfy
the differential equations 
\bea
\ddot x=-2\Omega \Pi _t\dot y\,,\qquad\ddot y=2\Omega \Pi _t\dot x\,,
\eea
with general solutions
\bea
x(\lambda) &=& A \cos\left(2\Omega \Pi_t \lambda\right) 
-B\sin\left(2\Omega \Pi_t \lambda\right)+x_0\,, \\
y(\lambda) &=& A \sin\left(2\Omega \Pi_t \lambda\right) +B\cos\left(2\Omega 
\Pi_t \lambda\right)+y_0\,.
\eea
Therefore, the projection of all geodesics on the (x,y) plane are circles with 
arbitrary origin $(x_0,y_0)$, and radius
\bea
r^2=A^2+B^2=\frac{\Pi_t^2-m^2-\Pi_3^2}{4\Omega^2 \Pi_t^2}\,.
\eea
All null ($m^2=0$) or timelike ($m^2>0$) geodesics project on circles in the 
$(x,y)$ plane with radii smaller than $1/2\Omega$. For $\Pi_3=0$, all null 
geodesics have $r_g=1/2\Omega$, independently of $\Pi_t$, and all spacelike 
($m^2<0$) geodesics project to circles with $r>r_g$. Furthermore, when $r>r_c$ 
the projections of these spacelike geodesics are closed timelike curves.

After presenting the geodesics in Cartesian coordinates, it is easy to make
contact with the discussion in polar coordinates of \cite{calvao, paiva}. As 
a consequence of homogeneity, it is obvious that all geodesics
with the same values of $\Pi_t, m^2, \Pi_3$ are just copies of a single one,
only shifted in the $(r,\phi)$ plane. For instance, the geodesics 
centered at the origin have angular momentum 
$L=-\left(\Pi_t^2-m^2-\Pi_3^2\right)/4\Omega \Pi_t$, 
and as we shift their center away from the origin, their angular momentum 
increases, becoming $L=0$ when they go through the origin. Furthermore, one 
can prove that as we shift the geodesics from being centered at the origin, 
they touch the chronologically safe region for $L=\Pi_t \left(1-
\sqrt{1-m^2/\Pi_t^2}\right)/\Omega $. This is the range of classical geodesics 
contained inside the chronologically safe region; note in particular that for
null geodesics ($m^2=0$), the geodesic with $L=0$, goes both through the 
origin and the boundary of the chronologically safe region (see figure 1).

It is worth noting that the $l^2=0$ metrics appear in supergravity 
solutions where other background fields are turned on \cite{gauntlett, 
horowitz}. This opens up the possibility of having particles charged under 
those fields, whose motion is determined by the combined effects of the metric 
and their coupling to the rest of the background. In this broader context, 
our discussion applies to neutral particles only, since we concentrate in 
solutions of General Relativity, with no further fields turned on. On the 
other hand, the geodesics of charged particles in supergravity backgrounds
have been discussed in \cite{rey}, and quite remarkably, it is possible to 
find causal charged geodesics with $L=0$ that leave the chronologically safe 
region.

\noindent
\paragraph{Hyperbolic plane $(l^2> 0)$:} The analysis of the geodesics is 
easiest in
hyperbolic coordinates. Taking advantage of the isometries of the metric,
we derive that the geodesics satisfy
\bea
\begin{gathered}
\dot t =\frac{\Omega^2-l^2}{l^2}\Pi_t-2\Omega Y \Pi_x\,,\\
\dot X = -2\Omega Y \Pi_t +4l^2Y^2\Pi_x\,,\\
\dot Y=4l^2Y\Pi_x(X-X_0)\,,\\
\dot X_3=\Pi_3\,,
\end{gathered}
\eea
where $\Pi_t,\Pi_x,\Pi_3$ are conserved quantities associated to the respective
isometries and $X_0$ is an arbitrary constant. From this and the ``on-shell'' 
condition it follows that the projections of the geodesics on the plane are 
circles
\bea
(X-X_0)^2+\left(Y-\frac{\Omega \Pi_t}{2l^2\Pi_x}\right)^2=
\frac{\Pi_t^2-m^2-\Pi_3^2}{4l^2\Pi_x^2}\,,
\eea
Let us concentrate on $m=\Pi_3=0$.
Compared to the $l^2=0$ case, there is a new 
feature. Since in hyperbolic 
coordinates $Y\geq 0$, the geodesic projects into a full circle only if the 
radius is smaller than $Y_0=\frac{\Omega \Pi_t}{2l^2\Pi_x}$. For null 
geodesics this condition implies 
$l<\Omega$, which precisely coincides with the range where there are 
closed timelike 
curves. So, if $l\leq \Omega$, all null or timelike geodesic project to full
circles, while only some of the spacelike geodesics do; among these, the ones
with $r>r_c$ project to the closed timelike curves. On the other hand, if we
consider $l>\Omega$, some timelike geodesics will still project to circles, 
but no null or spacelike geodesics will, and 
there are no closed timelike curves.
\EPSFIGURE{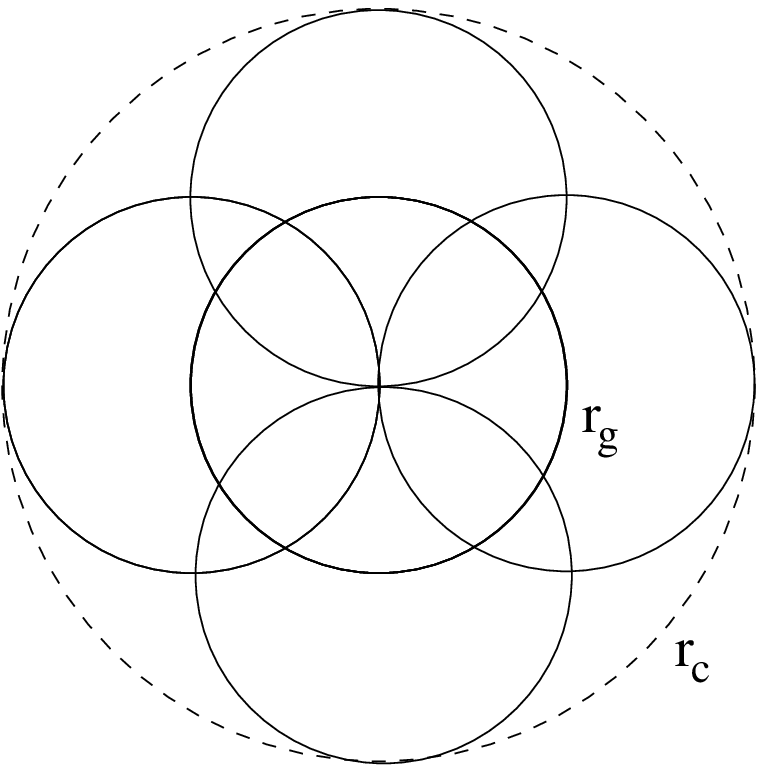, height=5cm}{The radius $r_c$ of the cavity is 
twice the radius of the null geodesic, $r_g$} 
In the Landau problem on $\bb{H}_2$, this can be phrased \cite{comtet} as 
saying that the magnetic field has to be strong enough, in order to force the 
electrons to follow closed orbits.

Going back to polar coordinates, one can show that the radius of the 
projection of a null geodesic satisfies  $\tanh 2lr_g=l/\Omega$, and
again we have the relation $r_g=r_c/2$.  This relation is easy to 
understand geometrically. The radius $r_g$
is that of null geodesics {\it centered} at a given point. On the other hand, 
the chronologically safe cavity is constructed by considering the family of 
null geodesics {\it going through} that same point. By homogeneity, all these
null geodesics, project to circles or radii $r_g$, so as illustrated in figure
1, they form a cavity or radius $r_c=2r_g$.

\section{The Klein-Gordon equation}

After analyzing the classical geodesics in these backgrounds, and comparing
them with the classical orbits for the Landau problem, we turn in this section
to the quantum version of the relation between G\"odel type metrics and the
Landau problem, by studying the solutions to the Klein Gordon equation in these
backgrounds, first considered in \cite{figue}. For the $l^2=0$ case,
the solutions to the massless Klein-Gordon equation were also discussed in
\cite {rey}. For the different signs of 
$l^2$, we will see many similarities, but also important differences, with 
the respective Landau problems. The reason for the
difference is that there is a coupling between energy and angular momentum for
probes in these backgrounds, so in the Landau analogy, the energy plays the 
role of effective charge (this was already observed in \cite{russo}).

\noindent
\paragraph{Flat plane $(l^2=0)$:} 
The Laplacian on this space is given in polar coordinates by
\bea
\Delta = \frac{1}{r}\partial_r(r\partial_r)
  +\left(\frac{1}{r}\partial_\phi-\Omega r\partial_t\right)^2
  -\partial_t^2\  +\partial _3^2\,.
\eea
Taking the ansatz
\bea
  \Psi = \Phi(r) e^{i\omega t +im\phi+ik_3x_3}\,, \qquad m\in\bZ\qquad w>0,
\eea
results in the following radial equation for a field of mass $M$
\begin{equation}
  \frac{1}{r}\partial_r(r\partial_r)\Phi
  -\left(\frac{m^2}{r^2} + \Omega^2\omega^2r^2\right)\Phi 
  + \left(\omega^2+2m\Omega\omega-M^2-k_3^2\right)\Phi=0\,.
\end{equation}
The $r$ dependent terms are (twice) the Schr\"odinger operator of a two 
dimensional harmonic 
oscillator with angular momentum $m$ and frequency 
$\hat{\omega}=\Omega\omega$. One can thus use the information regarding
the solution to the Schr\"odinger equation for a two dimensional
harmonic oscillator in polar coordinates. The wave functions can 
be written as
\begin{equation}
  \Phi_{n_r\,,m} (r) = C_{n_r\,,m}\,r^{|m|}\,e^{-\Omega\omega r^2/2}\,
  _1F_1(-n_r,\,|m|+1;\,\Omega\omega r^2 )\,.
 \label{eq:radialsol}
\end{equation}

If we require regularity of the wave functions at the origin and 
normalizability at infinity, it forces the first two arguments 
of the confluent series to be integers; in this case the confluent series 
truncates to a polynomial, which turns out to be proportional to the 
associated Laguerre polynomial. The boundary conditions would be modified 
if we take the domain wall \cite{we} into account, but we still expect this 
to be a good approximation. All in all, our wave functions can be written as 
follows:
\begin{equation}
  \Psi_{n_r\,,m} (t,\,\phi,\,r) = C_{n_r\,,m}\,r^{|m|}\,
  e^{-\Omega \omega r^2/2}\, L^{|m|}_{n_r} (\Omega \omega r^2 )\, 
  e^{im\phi}\,e^{i\omega t}e^{ik_3x_3}\,.
 \label{eq:wavesol}
\end{equation}
The energy is quantized, and it depends on the angular momentum $m$ and 
the radial quantum number $n_r$
\begin{equation}
  E = \hat{\omega}\left(|m| + 2n_r+1\right)\,, \qquad n_r = 0,1,\dots
 \label{eq:energy}
\end{equation}
The degeneracy at level $N\equiv |m|+2n_r$ is $N+1$.

The wavefunctions of \eqref{eq:wavesol} are the same as those of 
the single particle states of the QHE on
the plane. To make contact with the Landau problem, we define $n=(N-m)/2$, 
which equals $n_r$ for $m>0$ and $N-n_r$ for $m<0$. This quantum number $n$ is
the ordinary Landau level, and at each level $n$ we have infinite degeneracy, 
with $m=-n,\dots, \infty$. Note however that the previous wavefunctions have
a level dependent rescaling compared to the ordinary wavefunctions of the QHE, 
since they depend on $\Omega\omega r^2$, rather than just $r^2$. We comment 
below on the consequences of this rescaling. 

The frequency $w$ of a given level $n$ is determined by equating the constant 
term in the Laplacian with twice the energy eigenvalues of the 2d harmonic 
oscillator
\begin{equation}
  \omega^2+2m\Omega\omega-M^2-k_3^2 = 2\Omega\omega\,(N+1) \,,
\end{equation}
which determines the set of allowed frequencies to be
\begin{equation}
  w=(2n+1)\Omega \pm \sqrt{(2n+1)^2\Omega^2+M^2+k_3^2}\,, \hskip1cmn=0,1,\dots
 \label{eq:wspec}
\end{equation}

In what follows, we restrict to solutions with the positive sign of the square root, 
and focus on the sector $M^2=k^2_3=0$. The allowed frequencies are
\bea
w=2\Omega (2n+1), \hskip1cm n=0,1,\dots
\eea
and the levels are equally spaced. 

Let us switch to complex coordinates $z=re^{i\phi}$. 
The metric and Laplacian read
\bea
&ds^2=-\left(dt+\frac{i\Omega}{2}(zd\bar z-\bar zdz)\right)^2+dzd\bar z\,, & \\
&\Delta =\left(-1+\Omega ^2z\bar z\right)\partial ^2_t
+2i\Omega\left(\bar z\bar \partial-
z\partial\right)\partial _t +4\partial \bar \partial \,.&
\eea
The wavefunctions of the lowest level $(n=0)$ are given by
\bea
\Phi_{n=0,m}=z^me^{-\Omega^2 z\bar z}\,,
\eea
and except for the first ones, they are peaked outside the chronologically
safe region. We will also be interested in the states with the minimal value
of $m$ at each level $n$, i.e. the lowest weight states with $m=-n$. Their 
eigenfunctions are
\bea
\Phi _{n,m=-n}=\bar z^n e^{-\Omega ^2(2n+1)z\bar z}\,.
\eea
These are strongly peaked at
\bea
r=\sqrt{\frac{2n}{2n+1}}\frac{1}{2\Omega}\,,
\eea
so that these radii approach, for large $n$, the classical radius of null geodesics,
$r_g=1/2\Omega$. This is to be contrasted with the ordinary Landau problem
in the plane, where the lowest weight wavefunctions are strongly peaked at a 
radius that is unbounded as the level grows. The difference is caused by the 
fact that the energy of the state plays the role of an effective charge. As a result 
of the rescaling, all the lowest weight states fit inside the chronologically safe 
region.

The wave function of the harmonic oscillator is a polynomial of degree $N$ 
times the exponential $\exp -\Omega\omega r^2/2$. 
For a massless field this has a maximum at
\bea
r=\sqrt\frac{N}{2n+1}\frac{1}{\Omega}\,.
\eea
Beyond that radius, the wave function decays exponentially. Therefore the 
wavefunction is localized within the chronologically safe region as long 
as $N\lesssim2n+1$, or 
$m\lesssim0$. We saw the same phenomenon at the classical level, where all the 
null geodesics with angular momentum $L\leq0$ were fully within the radius 
$r_c$.

We conclude that wavefunctions with $m\leq 0$, are exponentially suppressed 
outside the chronologically safe region. On the other hand,
wavefunctions with $m>0$ may have a large support outside the 
radius of the chronologically safe region, $r_c$.
 
The $l^2=0$ solutions can be embedded in string theory, and the spectrum 
computed at weak coupling. This metric appears in \cite{horowitz} as 
a KK reduction 
of a 5d metric with
a compact direction; this model is exact to all orders in $\alpha '$. The 
spectrum was derived in the bosonic case in \cite{russo} and in the type II
and heterotic cases in \cite{russohet}. An important feature is that in the 
world-sheet, the vorticity $\Omega$ only couples to, say, the right-movers,
so only the right-moving zero modes appear in the Hamiltonian. The left-moving
zero modes don't appear neither in the Hamiltonian nor the level matching
conditions, so at every stringy level, there is infinite discrete degeneracy,
just as in the ordinary Landau problem in the plane.

\noindent
\paragraph{Sphere $(l^2<0)$:} With the standard change of coordinates $z=2R\tan\frac{\theta}{2}
e^{i\phi}$, and defining $K=\left(1+z\bar z/4R^2\right)^{-1}$, 
the metric and Laplacian 
read now
\bea 
&&ds^2=-\left(dt+i\frac{\Omega}{2}K(zd\bar z-\bar zdz)\right)^2
+K^2dz d\bar z\,,
\\
&&\Delta =(\Omega ^2z\bar z-1)\partial _t^2 +\frac{2i\Omega}{K}(\bar z
\bar \partial -z\partial)\partial _t +\frac{4}{K^2}\partial \bar \partial\,.
\eea
The allowed frequencies are
\bea
\omega =(2n+1)\Omega +\sqrt {(2n+1)^2\Omega^2+\frac {n(n+1)}{R^2}+M^2+k_3^2}\,,
\hskip1cm n=0,1,\dots
\eea
and for each frequency (Landau level), we have finite degeneracy $-n\leq m
\leq 4\Omega w(n)R^2$. At each level, the wavefunctions with the 
minimal value of $m$, i.e. the lowest weight states, are particularly easy
to find, and from them one can build in principle all the wavefunctions (see 
\cite{figue} for expressions). They are 
\bea
\Phi _{n,m=-n}
=\frac{\bar z^n}{\left(1+\frac {z\bar z}{4R^2}\right)^{2\omega (n) 
\Omega R^2+n}}\,.
\eea
For instance, if $M^2=k_3^2=0$, the lowest 
Landau level has $\omega =2\Omega$ and wavefunctions
\bea
\Phi_{n=0,m}=
\frac {z^m}{\left(1+\frac {z\bar z}{4R^2}\right)^{4\Omega ^2 R^2}}\,,
\hskip1cm m=0,\dots, 8\Omega^2 R^2\,.
\eea
The wavefunctions for lowest weight states are strongly peaked at 
\bea
\tan^2 \frac{\theta}{2}=\frac{n/4R^2}{\omega (n)\Omega+n/4R^2}\,,
\eea
which again in the large $n$ limit reproduce the classical value, $\cot
\theta _g=2\Omega R$.

This spectrum shares some common features with the Landau problem on the 
sphere (reviewed in the appendix): It has an infinite number of levels 
(labeled by $n$), and finite degeneracy for each of them. As in the $l^2=0$
case, a first difference is that at higher levels, there is a rescaling of
the wavefunctions, such that all the lowest weight states peak over the same
classical radius. On top of that, there is another 
important difference with the spectrum of the Landau problem on the sphere. In
that case, the curvature correction is quadratic in the level (compared to the
ordinary linear term), while here, for large $n$, the frequencies are linear 
in $n$.

\noindent
\paragraph{Hyperbolic plane $(l^2>0)$:} 
Finally, in this case we define $z=\frac{\tanh lr}{l} 
e^{i\phi}$, and the metric is identical to the $l^2<0$ case, but now 
with 
\bea
K=\frac {1}{1-l^2z\bar z}\,.
\eea
The frequencies are
\bea
\omega=(2n+1)\Omega +\sqrt {(2n+1)^2\Omega ^2-4l^2n(n+1)+M^2+k_3^2}\,.
\eea

Again, in the regime $l^2\leq \Omega ^2$, the frequencies are linear in $n$, 
for large $n$. The wavefunctions for the lowest weight states and the lowest 
level are very
similar to those of the $l^2<0$ case. For $M^2=k_3^2=0$ they are
\bea
\Phi_{n,m=-n}&=&\frac {\bar z^n}
{\left (1-l^2z\bar z\right )^{\frac {\omega \Omega}{2l^2}-n}}\,,
\\
\Phi_{n=0,m}&=&\frac {z^m}{\left (1-l^2z\bar z\right )^{\Omega^2/l^2}}\,,
\eea

As it happened in the discussion of the classical geodesics, here we have
to distinguish again between the $l<\Omega $ and $l\geq \Omega $ regimes. For 
$l^2< \Omega ^2$---the range where the metrics have CTCs---there is no upper
bound on $n$, and we have again an infinite number of levels. On the other 
hand, if we allow for $l^2>\Omega^2$, there is an upper bound on the level 
$n$, given by \cite{figue}
\bea
2n+1\leq \frac{\Omega}{l}\frac{\sqrt{M^2+k_3^2}}{\sqrt{l^2-\Omega^2}}\,,
\eea
and furthermore, above that level, states with a continuous energy spectrum 
appear. This matches the behavior
of classical geodesics discussed in section 2: for $l\leq \Omega$, all 
timelike or null geodesics project to closed orbits, and that is reflected at 
the quantum level by a discrete energy spectrum. On the other hand, for 
$l> \Omega$ some timelike geodesics and all null geodesics project to unbounded
orbits in the hyperbolic plane. At the quantum level, these correspond to
the states in the continuum of energy. This is completely analogous to 
the Landau problem on the hyperbolic plane \cite{comtet} (see also the 
appendix), where there is a finite number of discrete levels, and above 
it, a continuum of energy states. 

Note that at the point where the continuum appears, $l^2=\Omega^2$, the 
metric \eqref{eq:rebou} becomes $AdS_3\times \bR$. 
For string theory on $AdS_3$ the continuous representations of $sl(2,\bR)$ 
correspond to long strings \cite{michelson, seiberg}, as explained in 
\cite{malda}. However, the long strings can be understood as coming from
the spectral flow of spacelike geodesics \cite{malda}, while here the 
continuous representations correspond to massive particles (and therefore
timelike geodesics). It would be nice to understand better the relation, if
any, between these two appearances of $SL(2,\bR)$ continuous representations. 

\section{Holographic screens.}

As mentioned in the introduction, we don't expect the full G\"odel type 
solutions to be a valid background in string theory. On the other hand,
in previous work \cite{we}, we presented a 10d solution that patches a single
chronologically safe region to an outside metric, separated by a domain
wall made of supertubes. This solution is free of closed timelike curves,
and for an observer at the origin, Bousso's prescription associates a 
holographic screen with the same radius as in the original solution. This 
solution with a domain wall allows us to address the existence of a holographic
description for a region of the universe that is locally G\"odel type, freeing
us from all the conceptual troubles associated with closed timelike curves.
This holographic screen is presumed to encode the physics localized
inside the domain wall, although the precise meaning of ``localized'' in this
context is not completely clear, since the region on the inside of the domain 
wall is not causally disconnected from the rest.

In the Landau problem analogy, restricting to a single chronologically safe
region, bears some resemblance to
considering a Hall droplet of finite size \cite{halperin}, instead of the 
Quantum Hall effect in the full plane. For instance, in the 
matrix quantum mechanics description of the QHE for a finite size droplet,
due to Polychronakos \cite{polyc}, one introduces a harmonic potential to 
confine the electrons into a finite domain, breaking the translation 
invariance of the system. In the G\"odel type solutions, the domain wall has 
a similar effect.

Bousso's prescription for constructing holographic screens is completely
classical. As such, there is no guarantee that given a classical solution
of general relativity, there is a quantum gravity version of that background 
and much less a holographic description. In what follows, we assume that such 
a holographic description exists, for a single chronologically safe region of 
the family of 4d metrics we have been studying, and present some heuristic 
ideas on what this holographic theory might be. Eventually, it is plausible 
that these ideas will only make sense when these metrics (or rather, metrics 
with domain walls surrounding a single chronologically safe region) are 
embedded in solutions of string/M theory.

First we address the question of which bulk states we expect inside the domain
wall. Properly, we should impose boundary conditions consistent with the
presence of a domain wall, but we believe that for states inside the cavity,
the analysis of the previous section is a good approximation. A natural 
prescription is 
to restrict to states whose wavefunctions start
decaying exponentially inside the chronologically safe region, or just at the
boundary. For instance, in the previous section we saw that for massless 
states, those wavefunctions have arbitrary $n$ and $m\leq 0$ (they correspond
to null geodesics with $L\leq 0$, which are fully inside the chronologically
safe region).

Although the presence of the domain wall naturally suggests a way to 
restrict the bulk states, a pure QFT analysis misses the effects of gravity. 
Once we consider the backreation of these states to the metric, we expect a 
cut-off in the allowed frequencies: the Klein Gordon wavefunctions are 
extended in $x_3$, and localized in the plane of constant $t,x_3$, so once the 
backreaction is taken into account, the new solution will be similar to a 
cosmic string. In particular, it will introduce a deficit angle, and we should 
require that this deficit angle is smaller than $2\pi$, i.e. we 
require that $w$ is smaller than the mass of a particle that would close the
universe in 2+1 dimensions, $w\leq 1/\ell _P$. This translates into 
$n_{max}\sim r_c/\ell_P$.

Having discussed the spectrum of modes we expect inside the domain 
wall, we address next the boundary degrees of freedom. First, let's recall 
where the holographic screen is, 
according to Bousso's prescription. If we restrict 
to the non-trivial 3d part of the metrics, the position of the holographic
screen was computed in \cite{horava}, and it is best described in polar 
coordinates. By the symmetries of the problem, it was
argued in \cite{horava} that the radius of the screen is determined by 
$dg_{\phi\phi}(r)/dr=0$, or
\bea
\sinh lr_h=\frac{1}{\sqrt{2}{\sqrt {\frac{\Omega ^2}
{l^2}-1}}}\,.
\eea
This is always smaller than the radius of the cavity $r_c$ and bigger than
the radius of the null geodesics, $r_g<r_h<r_c$. In fact,
\bea
\sinh lr_h=\frac{1}{\sqrt 2} \sinh lr_c\,.
\eea
For the full G\"odel type solutions, the holographic screen breaks the 
isometry of the full solution, as happens in other solutions with observer
dependent holographic screens ($e.g.$ , the static observer approach to 
holography for de Sitter space \cite{minic}). In this case, we have $\bR\times 
U(1)$ associated to time translations times rotations on the plane (plus an 
additional $\bR$ if we consider translations in the $x_3$ direction). On the
other hand, if we start with a solution with a domain wall, the holographic
screen breaks no further symmetries.

Next, we are going to relate the number of boundary degrees of freedom
with the Landau level structure of the bulk spectrum. In general, the number 
of degrees of freedom in the holographic screen scales 
with the area $A$ of the screen, $N\sim A/\ell_P^2$, and since the screen 
extends indefinitely in the $x_3$ direction, the area---and the number of 
degrees of freedom---is infinite. We instead consider a screen with a cut-off 
length $h$ in its height, and concern ourselves on how the different quantities
scale with $r_c$, for fixed $h$. The area goes like $A\sim r_c h$ and
$N\sim r_c h/\ell_P^2$, so the expected number of degrees of freedom (for 
fixed $h$) grows linearly with $r_c$.

Now we use this relation $N\sim r_c$ and $n_{max}\sim r_c/\ell_P$ to conclude
that $N_{dof}\sim n_{max}$. Namely, we roughly get a boundary
degree of freedom per level per field\footnote{Although in the previous
section we only discussed the Klein Gordon equation for a scalar field, in 
string theory realization of the $l^2=0$ case, all fields (graviton, NS form) 
also display a Landau level type
spectrum \cite{russo}.}. This can be realized at least in a couple of
ways: The independent degrees of freedom map to the states with $n=n_{max}$, or
alternatively, for each level there is an independent boundary degree of 
freedom. 

We would like to conclude by presenting a simple model of a holographic 
description that captures this scaling, although clearly leaves many things 
out. Simply put, we want to substitute the classical holographic screen by a
fuzzy version of it, a non-commutative cylinder in the case at hand, with the 
scale of non-commutativity given by the Planck area, $\ell_P^2$. Our chief 
motivation is that in general, when the classical screens given by Bousso's 
prescription have finite proper area, the covariant version of the holographic 
principle assigns them a finite number of degrees of freedom at the quantum 
level, and a natural way to assign a finite number of degrees of freedom to 
a surface is by considering a matrix regularization. Similar ideas have been 
proposed in \cite{nastia, tom}, in the context of the static  observer 
approach to holography in de Sitter space \cite{minic}, by replacing the 
classical spherical horizon with a fuzzy sphere.

In the context of the matrix model for M-theory, non-commutative
cylinders were constructed in \cite{bak} providing a matrix realization of
supertubes. Recall that in the 10d solution of \cite{we}, the domain wall 
separating the G\"odel type region from the exterior metric is made out of 
supertubes; however, the matrix theory of \cite{bak} doesn't quite describe
this domain wall, since the latter was constructed with {\it smeared} 
supertubes. It would be interesting to construct the matrix model of the
IIA solution of \cite{we}, with a $l^2=0$ G\"odel type region inside the
supertube domain wall; alternatively, one could consider the matrix model 
formulation for the full $l^2=0$ solution, and try to extract from it the 
physics of a single chronologically safe region.

The non-commutative cylinder is a solution of the matrix regularized version 
of membrane theory, with three time independent non-zero matrices satisfying
\bea
[X_1,X_2]=0\,, \hskip1cm [X_1,X_3]=-iaX_2\,, \hskip1cm [X_2,X_3]=iaX_1\,,
\eea
where $a$ is a length scale. Moreover, $X_1^2+X_2^2$ is a Casimir of 
this algebra, and we consider representations with a fixed value $X_1^2+X_2^2=
R^2$. These 
matrices correspond to a non-commutative cylinder of radius $R$ and 
non-commutativity scale $a$ in the $x_3$ direction. The quantum of area of this
non-commutative cylinder is $2\pi aR$, and we set $aR\sim \ell ^2_P$. Note that
the two sides of this quantum of area are very different, if $R \gg \ell _P$.

As already noticed in \cite{bak}, the scalar fluctuations around this matrix
configuration are also similar to the Landau problem on the plane. This is
intuitively clear, since a supertube is supported against collapse by an
electric field along the tube, and a magnetic field transverse to it. The
equation that describes fluctuations of a massless scalar is \cite{bak}
\bea
\left(\partial _t-\frac{a}{\ell^2_P}\partial _\theta\right)^2\Phi 
-\frac{a^2}{\ell_P^4}\partial _\theta ^2\Phi
-\frac{a^2}{\ell_P^4}R^2\partial _3^2\Phi =0\,.
\eea
The ansatz $\Phi=e^{iwt}e^{im\theta}e^{ik_3x_3}$ yields the frequencies
\bea
w=m\frac{a}{\ell_P^2}+ 
\sqrt{\left(m\frac{a}{\ell_P^2}\right)^2+\frac{a^2R^2}{\ell _P^4}
k_3^2}\,,
\eea
which reproduce the spectrum of frequencies for $l^2=0$, \eqref{eq:wspec}, if 
we identify
\bea
a/\ell_P^2 \; \leftrightarrow \, \Omega\,.
\eea
The difference is that now the spectrum has 
no degeneracy for a given frequency, which agrees with the
scaling we deduced previously, roughly one boundary degree of freedom per
Landau level. Note however, that at this level of discussion, we don't see
the existence of a cut-off, $n_{max}$.

Using $aR\sim \ell_P^2$, the previous identification leads to 
$R\sim 1/\Omega$. These preliminary 
considerations are clearly not enough to reproduce Bousso's prescription for 
the position of the holographic screen in the $l^2=0$ case, $r_h=1/\sqrt{2}
\Omega$.

To sum up, in the full homogeneous G\"odel type solutions, the Klein 
Gordon equation gives a degeneracy of solutions at each frequency, 
given by the quantization of momentum, characteristic of the Landau problem. 
The states of a given frequency form full representations of $SU(2)$, $H_2$
or $SL(2,\bR)$. Once we restrict to a single chronologically safe region, it
is natural to consider a truncation of the spectrum at each level, keeping 
states that are localized inside the region. These states clearly no longer 
form full representations of the corresponding groups. So far, everything 
we said refers to quantum field theory in these backgrounds. 

Finally, on a more speculative note, we tried to relate the boundary
degrees of freedom of a possible dual description with the states that are
localized inside a single chronologically safe region.

We seem to be a long way from understanding holography when the holographic
screens are not at asymptotia, assuming such a thing makes sense. Making more
precise the speculations presented in this last section will hopefully 
provide some of the much needed insight.

\section{Discussion}

We have pointed out the close relation between a family of 3+1 G\"odel type
solutions and the Landau problem of a charged particle moving on a surface,
in the presence of a constant magnetic field, a relation discussed 
in a particular case in \cite{rey}. This analogy has allowed us to suggest a 
heuristic picture of a potential holographic description for a single
chronologically safe region.

Our discussion generalizes immediately to other solutions. For instance,
the 5d G\"odel solution found in \cite{gauntlett} would be related to a
charged particle moving in $\bR^4$ with magnetic fields transverse to two
planes. This system was considered recently in \cite{polchinski}. Going in
the opposite direction, a system that has received a great amount of interest 
is a 4+1d generalization of the QHE \cite{zhang}. In light of our discussion, 
we can ask whether there is any (higher dimensional) GR solution related to 
this model. Since the 4+1d QHE is based on the second Hopf map (in the same 
way the ordinary QHE is based on the first Hopf map), natural candidates are
metrics of $SU(2)$ bundles over $S^4$, but to obtain a Lorentzian signature, 
it seems more natural to consider the reformulation of this higher dimensional
QHE as a $U(1)$ fibration over $\bC\bb{P}^3$ \cite{karab} \footnote{We would 
like to thank D. Karabali and V.P. Nair for pointing out this 
possibility.}.

Our suggestions about a holographic description of these backgrounds are far
from conclusive. Another approach to holography for general backgrounds, 
has been suggested by Banks \cite{banks}, who put forward the idea that 
there is an approximation to quantum gravity, dubbed {\it asymptotic 
darkness}, where the possible black 
holes in that background are stable, and the microstates of all possible black 
holes in that background provide a basis of the Hilbert space of the quantum 
theory. In the case of the supersymmetric 5d G\"odel solution 
\cite{gauntlett}, there has been some work on the possible black 
holes \cite{eric}. On the other hand, one might hope to identify 
the string states that become black holes, once the interactions are taken 
into account. This would require making precise the bound on frequencies 
mentioned in section 4. The 5d G\"odel solution (or more properly, a single 
chronologically safe cavity of it) might be a candidate to make more concrete 
the idea of asymptotic darkness.

An important point is that in the present work we considered only
solutions to General Relativity. In Bousso's prescription, the position of the 
holographic screen is determined by the metric alone, or in other words, by 
the geodesics of neutral particles. Once we consider solutions in extended 
theories, like supergravity, we face the possibility of having solutions with 
the same metric but differing in the rest of the fields turned on, and Bousso's
prescription is insensitive to that difference. Since the holographic 
principle refers only to the {\it number} of degrees of freedom, but not their
{\it nature}, we see not immediate contradiction: the putative holographic 
descriptions of these backgrounds could be two different theories, with the 
same number of degrees of freedom. On the other hand, as the 
results of \cite{rey} nicely illustrate, the geodesics of charged objects can
have dramatically new behavior compared to the neutral ones, and it seems
important to understand their relevance for holography in spacetimes with
closed timelike curves \cite{horava}, and the holographic principle 
in general.

In the present work, we suggested that if there is a holographic description
of a single chronologically safe region for G\"odel type universes, it will
involve the quantum mechanics of the matrix regularization of the classical 
holographic screen, $i.e.$ a non-commutative cylinder. The obvious 
generalization of this idea is that whenever Bousso's prescription yields a 
classical holographic screen with finite proper area, the quantum holographic 
description involves a matrix regularization of the classical screen. 

\bigskip

\subsection*{Acknowledgments} 
We would like to thank Ofer Aharony, Micha Berkooz,
Raphael Bousso, Jorge Russo and especially Tom Banks and Gary Gibbons for
useful discussions. BF would like to thank the Institute for Particle Physics
at the University of California at Santa Cruz for hospitality at early stages
of this project. ND and BF would like to thank the organizers of the Benasque
workshop on string theory for inviting them, and for a stimulating atmosphere.
JS would like to thank the Aspen Center for Physics for hospitality during
the last stages of this project. BF is supported through a European 
Community Marie Curie Fellowship, and also by the Israel-U.S. Binational 
Science Foundation, the IRF Centers of Excellence program, the European RTN 
network HPRN-CT-2000-00122 and by Minerva. JS was partially supported
by the Phil Zaccharia fellowship and is currently supported by the United
States Department of Energy grant number DE-FG02-95ER40893.

\appendix
\section {The Landau problem in surfaces of constant curvature.}

In this appendix we review the Landau problem of a charged particle moving 
under the presence of a constant magnetic field, perpendicular to a surface
of constant curvature, following \cite{comtet, dunne}. We consider the sphere, 
the plane, and the hyperbolic
plane. The spectrum is related to representations of $SU(2)$, $H_2$ and 
$SL(2,\bR)$ respectively. 

In each case, we will have a set of discrete energy levels. For the sphere, 
the degeneracy of each level is finite, while for the plane and the hyperbolic
plane, each level has infinite degeneracy. On the other hand, the number of
discrete levels is infinite for the sphere and the plane, but finite for the
hyperbolic plane. Finally, in the hyperbolic plane, 
there is a continuum of states, above the finite range of discrete levels.

We have a surface with metric
\bea
ds^2=g_{z\bar z}dz d\bar z\,.
\eea
The volume form is 
\bea
dv= \frac {ig_{z\bar z}}{2}dz\wedge d\bar z\,.
\eea
The natural definition of constant magnetic field is $F=Bdv$, and the
Hamiltonian for a charged particle moving in this surface is
\bea
H=\frac {1}{\sqrt {g}}(P-A)g^{z\bar z}\sqrt {g}(P-A)\,.
\eea
We start considering the 2d surface to be $S^2$. In this case, $N\equiv 2BR^2$ 
has to be an integer, $g_{z\bar z}=1/\left(1+z\bar z/4R^2\right)^2$, and with the gauge 
choice
\bea
A_z=\frac{-iB}{4}\frac{\bar z}{1+\frac{z\bar z}{4R^2}}\,,
\eea
the Hamiltonian reads
\bea
H=\frac{1}{2R^2}\left(-{1+\frac{z\bar z}{4R^2}}^2\partial \bar \partial 
+\frac{B}{4}\left(1+\frac{z\bar z}{4R^2}\right) (\bar z\bar \partial
- z \partial)+\frac{B^2}{16}z\bar z\right)\,.
\eea
We introduce
\bea
\begin{aligned}
L_+&=-\frac{1}{2R}z^2\partial -2R\bar \partial +\frac {BR}{2}z\,,
\\
L_-&=\frac{1}{2R}\bar z^2\bar \partial +2R \partial +\frac {BR}{2}\bar z\,,
\\
L_3&=z\partial-\bar z\bar \partial -BR^2\,,
\end{aligned}
\eea
which form an $SU(2)$ algebra. In terms of these operators, the Hamiltonian
can be rewritten as
\bea
H=\frac{1}{2R^2}\left(\frac{1}{2}(L_+L_-+L_-L_+)+L_3^2-B^2R^4\right)=
\frac{1}{2R^2}\left (C_2-B^2R^4\right)\,,
\eea
where $C_2$ is the second Casimir of $SU(2)$. This allows for a purely 
algebraic deduction of the spectrum. The energy spectrum is 
\bea
E_n=B\left(n+\frac{1}{2}\right)+\frac {n(n+1)}{2R^2}\,.
\eea
There is the familiar linear term, plus a quadratic correction due to the
curvature. The unnormalized lowest Landau level wavefunctions are
\bea
\Phi_{n=0,m}=\frac {z^m}{\left(1+\frac{z\bar z}{4R^2}\right)^{\frac{N}{2}}}\,,
\eea
i.e. the wavefunctions
are the wavefunction of the vacuum times a holomorphic polynomial. Each Landau 
level is finite dimensional, the degeneracy of the $n$-th level being 
$N+2n+1$, the states are in SU(2) representations. The generic wavefunctions
can be obtained by applying $L_+$ on the lowest weight states of each level, 
defined by $L_-\Phi=0$. The lowest weight wavefunctions are
\bea
\Phi _{n, m=-n}
=\frac{\bar z^n}{\left(1+\frac{z\bar z}{4R^2}\right)^{\frac{N}{2}+n}}\,.
\eea
Having discussed the Landau problem on the sphere, one can recover the Landau
problem on the plane as a limit $N,R\rightarrow \infty$, keeping $B=N/2R^2$
fixed. By an appropriate rescaling of the operators, the $SU(2)$ algebra 
contracts to a Heisenberg algebra. The curvature correction term drops from 
the energy spectrum, and we are left with the familiar
\bea
E_n=B\left(n+\frac{1}{2}\right)\,.
\eea
Now each Landau level has infinite degeneracy (since we sent $N\rightarrow
\infty$), and the lowest Landau level
and lowest weight wavefunctions are just the limit of the ones on the sphere
\bea
\Phi_{n=0,m} &=& z^me^{-\frac{Bz\bar z}{4}}\,,
\\
\Phi_{n,m=-n} &=& \bar z^me^{-\frac{Bz\bar z}{4}}\,.
\eea
Finally, we can consider the Landau problem in the hyperbolic plane. Now
$g_{z\bar z}=1/\left(1-z\bar z/4R^2\right)^2$, 
and most of the previous discussion
goes through. There is in principle no quantization on $B$, the $SU(2)$
is replaced by an $SL(2,\bR)$ algebra and the discrete spectrum is
\bea
E_n=B\left(n+\frac{1}{2}\right)-\frac {n(n+1)}{2R^2} \,,
\hskip1cm{0\leq n< \big \lfloor BR^2\big \rfloor -1}\,.
\eea
There is only a finite number of discrete Landau levels, each with infinite 
degeneracy, with states forming discrete $SL(2,\bR)$ representations. Beyond
that range of energies, there is a continuous part to the 
spectrum \cite{comtet}, with states in the continuous principal series 
${\cal C}_j$
\bea
&&j=-\frac{1}{2}+i\nu\,, \hskip1cm 0\leq \nu \leq \infty\,,
\\
&&E(\nu)=\frac{1}{2R^2}\left(\frac{1}{4}+BR^2+\nu^2\right)\,.
\eea
Note that $E_n< \left(1/4+B^2R^4\right)/2R^2$, while 
$E(\nu)\geq \left(1/4+B^2R^4\right)/2R^2$, so
the discrete and continuous spectra don't overlap.


\end{document}